\documentclass[prl,english,preprintnumbers,amsmath,amssymb,nofootinbib,twocolumn,superscriptaddress]{revtex4-1}

\pdfoutput=1

\usepackage[latin1]{inputenc}
\usepackage{graphicx}
\usepackage{bbm}
\usepackage{amssymb}
\usepackage{amsmath}

\usepackage{dsfont}

\def\0#1#2{\frac{#1}{#2}}

\def\s0#1#2{\mbox{\small{$ \frac{#1}{#2} $}}}

\newcommand{\Tr}{\mathrm{Tr}}

\newcommand{\E}{\mathrm{e}}

\newcommand{\be}{\begin{eqnarray}}
\newcommand{\ee}{\end{eqnarray}}

\newcommand{\nn}{\nonumber }

\newcommand{\zgc}{\mathcal Z}

\usepackage{babel}
\makeatother
\begin{document}

\title{Inhomogeneous phases in one-dimensional mass- and spin-imbalanced Fermi gases}

\author{Dietrich Roscher}
\affiliation{Institut f\"ur Kernphysik, Technische Universit\"at Darmstadt, 
64289 Darmstadt, Germany}
\author{Jens Braun} 
\affiliation{Institut f\"ur Kernphysik, Technische Universit\"at Darmstadt, 
64289 Darmstadt, Germany}
\affiliation{ExtreMe Matter Institute EMMI, GSI, Helmholtzzentrum f\"ur Schwerionenforschung GmbH,
64291 Darmstadt, Germany}
\author{Joaqu\'{\i}n E. Drut} 
\affiliation{Institut f\"ur Kernphysik, Technische Universit\"at Darmstadt, 64289 Darmstadt, Germany}
\affiliation{ExtreMe Matter Institute EMMI, GSI, Helmholtzzentrum f\"ur Schwerionenforschung GmbH,
64291 Darmstadt, Germany}
\affiliation{Department of Physics and Astronomy, University of North Carolina, Chapel Hill, NC 27599, USA}

\begin{abstract}
We compute the phase diagram of strongly interacting fermions in one dimension
at finite temperature, with mass and spin imbalance. 
By including the possibility of the existence of a spatially inhomogeneous ground state,
we find regions where spatially varying superfluid phases are favored over homogeneous phases. 
We obtain estimates for critical values of the temperature, mass and spin imbalance, above which these phases
disappear. Finally, we show that an intriguing relation exists between the general structure of the phase diagram and
the binding energies of the underlying two-body bound-state problem. 
\end{abstract}

\maketitle

{\it Introduction -} Continuous progress in ultracold-atom experiments in the last decade have made it possible to study 
strongly coupled quantum gases with an unprecedented degree of versatility~\cite{RevTheory,RevExp} and precision~\cite{PrecisionExp}. 
From the time of the first experiments able to cool down fermions into the degeneracy regime, multiple groups around the world have pursued the 
determination of the thermal and structural properties of these gases, with an increasing degree of control on parameters such as temperature, 
polarization, trapping potential, and interaction strength. Such advances have enabled, in addition, the study of mixtures of atoms with different 
masses, yet another variable in the multidimensional phase diagram of these remarkable systems. 

One of the most salient features of these degenerate Fermi gases in three spatial dimensions is that, at low enough
temperature, they may spontaneously break an internal symmetry to form a condensate. Depending on the value of the 
parameter~$k^{}_{\rm F} a^{}_{\rm s}$ (where the Fermi momentum~$k_{\rm F}$ is determined by the density $n$, and $a^{}_{\rm s}$ is the s-wave scattering 
length), the condensate may be a simple Bose-Einstein condensate (BEC) of di-fermion molecules (for strong interatomic attraction), or the well-known condensate of
Cooper pairs (in the weakly interacting BCS regime).
As $k^{}_{\rm F} a^{}_{\rm s}$ is varied along this BCS-BEC crossover, non-zero spin polarization and mass asymmetry may 
lead to exotic phases such as a Fulde-Ferrell-Larkin-Ovchinnikov (FFLO) phase~\cite{FFLO}. In the latter, 
the condensate displays spatial oscillations with a characteristic wavelength, before it is expected to disappear at large 
polarization or asymmetry.

In fact, several theoretical studies in three spatial dimensions have suggested the existence of phases of 
the FFLO type~\cite{inh3dref} (see also Ref.~\cite{inhreview} for a review). 
However, their experimental verification appears to be 
very challenging as it may only exist in a narrow band between the conventional BCS-type phase
in the core of the trap and a non-superfluid mixed phase in the outer layer. 
On the other hand, experiments in tightly constraining external potentials have also analyzed the problem 
in one spatial dimension~\cite{Moritz,Liao,Guan}, where it is expected that the phase diagram is to a large 
extent occupied by a phase of the FFLO type~\cite{Yang,Orso,Liu,Parish,Casula}. 

Such one-dimensional problems have attracted much interest in the last three decades, as bosonization 
and {density matrix renormalization group} techniques have been able to provide remarkable insights into the nature of the low-energy excitations, namely 
the Luttinger liquid (see, e.g. Ref.~\cite{Luttinger} for an introduction). Besides their relevance for condensed-matter systems, studies of 
inhomogeneous phases have recently gathered attention in other areas, such as the theory of the strong interaction~\cite{Kojo:2009ha}. 
The latter were triggered by Thies' pioneering analytic studies of inhomogeneous phases in one-dimensional relativistic fermion
models~\cite{Thies:2003br}.

In the present work, we provide a first study of the full finite-temperature phase diagram of strongly coupled, mass- and spin-imbalanced 
Fermi gases in one spatial dimension. In particular, we include an analysis of the fate of FFLO-type phases under a variation of the experimentally 
accessible parameters, such as the temperature and the mass difference of the fermions. To this end, we employ a mean-field approach, which can be viewed as the 
lowest-order approximation of the underlying path integral. Such mean-field studies of one-dimensional systems are usually expected to 
be at best qualitative. As is well known, long-range fluctuations hinder the spontaneous breakdown of a continuous symmetry at finite temperature in 
dimensions lower than three (in particular the breaking of the U($1$) symmetry associated with superfluidity). 
In current experiments, however, quasi one-dimensional systems are realized by a two-dimensional lattice array of atomic 
tubes~\cite{Moritz,Liao}. If the inter-tube coupling is weak but finite, a mean-field analysis of the system is justified, as explained in
Refs.~\cite{Yang, Parish}. In the following, we shall work in the strict one-dimensional limit, i.e. in the limit of vanishing transverse coupling, 
where the validity of mean-field theory may again be debatable~\cite{Zhao}. However, we assume that at least the qualitative features of our study
can be extrapolated to the regime with finite but weak transverse coupling.

 {\it Formalism -} In our analysis we consider a Hamiltonian that describes the dynamics of a theory with two
fermion species, denoted by \mbox{$\uparrow$ and $\downarrow$}, interacting only
via a zero-range two-body interaction:
\be
\hat{H}= 
-\int d^{}x \left (
 \sum_{\sigma=\uparrow,\downarrow}
  \hat{\psi}_{\sigma}^{\dagger}({x})\frac{\partial_x^2}{2m_{\sigma}}\hat{\psi}_{\sigma} ({x})
  + {g} \hat{\rho}_\uparrow ({x}) \hat{\rho}_\downarrow ({x}) \right )\,.\nn
\ee
The operators~$\hat{\rho}^{}_{\uparrow,\downarrow}$ are the particle density operators associated with the two 
fermion species ($\uparrow,\downarrow$). 
The coupling~$g$ is related to the s-wave scattering 
length~$a_{\rm s}$ by~$g ={1}/{a_{\rm s}}$.
For our analysis, we shall employ a path-integral representation of~$\zgc$:
\be
\zgc = \int {\mathcal D}\psi^{\dagger}{\mathcal D}\psi\, \E^{-{S} [\psi^{\dagger},\psi]
 }\,,\nn
\ee
{where the fields $\psi^{\rm T}=(\psi^{}_{\uparrow},\psi^{}_{\downarrow})$
obey anti-periodic boundary conditions in the 
imaginary-time direction, and}
\be
&& \!\!\!\!\! S [\psi^{\dagger},\psi]=  \int_0^\beta d\tau \int d x\, \bigg\{ \psi^{\dagger}
\left( \partial _{\tau} - \frac{1}{m_{+}}\partial_x^{\,2} -\bar{\mu} \right) \psi \nn\\
&& \quad -  \psi^{\ast}_{\uparrow}\left(\frac{1}{m_{-}}\partial_x^{\,2}+h\right)\psi^{}_{\uparrow} 
{+  \psi^{\ast}_{\downarrow}\left(\frac{1}{m_{-}}\partial_x^{\,2}+h\right)\psi^{}_{\downarrow}}
\nn\\
&& \qquad\qquad\qquad\qquad\qquad\qquad\quad\;\;\; + \, \frac{g}{2}(\psi^{\dagger}\psi) (\psi^{\dagger}\psi)
\bigg\}
\,\label{eq:action}
\ee
{with $\beta=1/T$ being the inverse temperature}. 
For convenience, we have introduced the average chemical potential~${\mu}=(\mu^{}_{\uparrow}+\mu^{}_{\downarrow})/2$ 
and the spin polarization parameter~$h=(\mu^{}_{\uparrow}-\mu^{}_{\downarrow})/2$. Moreover, we introduce 
simple measures for the mass imbalance:
\be
m_{\pm}={\frac{4 m_{\uparrow}m_{\downarrow}}{m_{\downarrow}\pm m_{\uparrow}}}\,,\quad
\bar{m}=\frac{m_{+}}{m_{-}}\,,
\quad m_{\rm r}={\frac{m_{+}}{4}}\,.
\ee
where~$0\leq |\bar{m}| < 1$ and $m_{\rm r}$ is the reduced mass. In the following we choose units such that $m_{+}=1$.

To derive the effective potential for the desired order parameter, which in this case corresponds 
to off-diagonal long-range order, i.e. condensation of Cooper pairs, we employ a Hubbard-Stratonovich 
transformation and introduce an {auxiliary bosonic field~$\varphi$ to} remove effectively the four-fermion interaction 
from the action. The fermions can then be integrated out {straightforwardly} to obtain the quantum effective action
~$\Gamma\sim -\ln {\mathcal Z}$:
\be
\label{eq:orderpot}
&&\!\!\!\!\!\Gamma[\varphi,\varphi^*] \nn\\
&& = -\beta \int dx \frac{|\varphi(x)|^2}{g} +\! \sum_{j=-\infty}^{\infty} \Tr \ln \left (
\begin{array}{cc}
K_j^{-} & \varphi(x) \\
\varphi^*(x) & K_j^{+} 
\end{array}
\right )\,.
\ee
Here, the sum covers the Matsubara frequencies~$\omega_j=(2j+1)\pi/\beta$, and $K^{\pm}_j=(-{\rm i}\omega_j \pm \partial_x^2 - \bar{m}\partial_x^2 \pm \bar{\mu} - h)$.
The ground state of the theory is then obtained by minimizing~$\Gamma$ with respect to the scalar fields. As we allow
for a spatially varying~$\varphi(x)$, we are able to detect inhomogeneous phases whenever they are energetically favored over their 
homogeneous counterparts.  

The minimization of the effective action is done numerically, and is based on two types of expansions: first, we 
write the effective action as sum of~$n$-point functions~$\Gamma^{(n)}$: $\Gamma = \sum_{n=0}^{n_{\text{max}}} \Gamma^{(2n)}\cdot(\varphi^*\varphi)^n$, where an integration over
the spatial coordinate is implicitly assumed. Second, we expand the field~$\varphi$ in plane waves, i.e.
\be
\varphi(x) = \alpha_0 + \sum_{l=1}^{l_{\text{max}}} \alpha_l \cos(l\omega_{\varphi} x)\,,\label{eq:phiexp}
\ee
which is motivated by standard FFLO theory~\cite{FFLO}. 
In practice, it is necessary to truncate both sums at some given values~$n_{\text{max}}$ and~$l_{\text{max}}$. The
stability of the results for the phase structure can then be tested by varying~$n_{\text{max}}$ and~$l_{\text{max}}$ (see also below). 
{The parameter~$\omega_{\varphi}$ has been used to}
optimize the expansion~\eqref{eq:phiexp} of the field~$\varphi$. More specifically, for given~$n_{\text{max}}$ and~$l_{\text{max}}$, we 
varied the parameters~$\{\alpha_l\}$ and~$\omega_{\varphi}$ to find the ground state (gs) of the theory.
{This state $\varphi_{\rm gs}(x)$ is then directly related to the 
fermion gap~$\Delta(x) \sim |\varphi_{\rm gs}(x)|^2$.}
\begin{figure}[t]
\includegraphics[width=1\columnwidth]{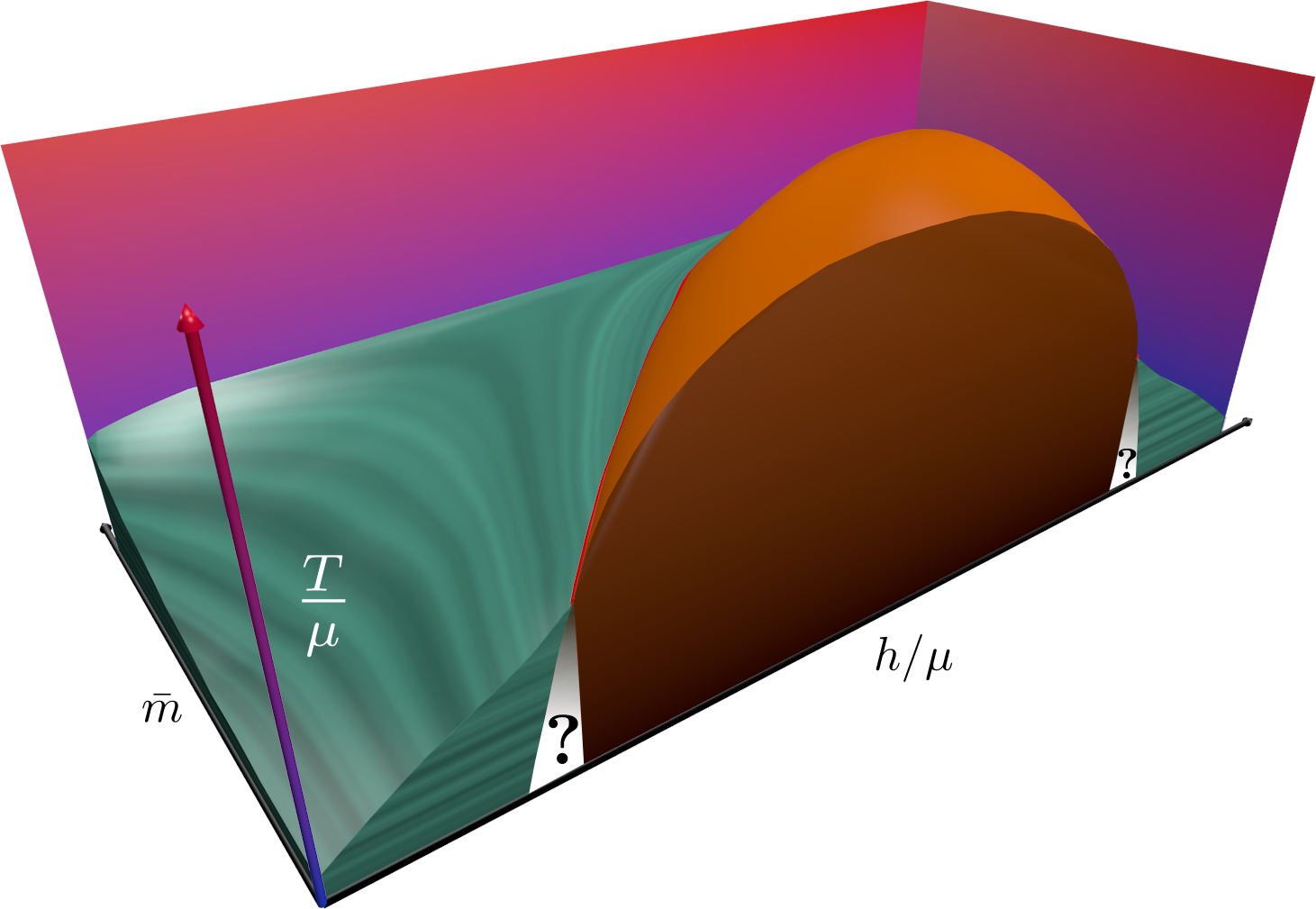}
\caption{\label{fig:p3d} 
(Color online) Phase diagram in the ($T$,$\bar{m}$,$h$) {space, where $0 \leq T/\mu \leq 0.6$, $0\leq \bar{m} \leq 0.95$
and~$-1\leq h/\mu \leq 1$,}
The (orange-brown) `dome' corresponds to the conventional 
BCS phase governed by a homogeneous ground state. The greenish-colored domains attached to the left and to the right of the BCS `dome'
represent the FFLO-type phase.
}
\end{figure}

{\it Results -} 
To verify our approach, we first computed the phase diagram of the {one-dimensional massless Gross-Neveu (GN) model}
in the plane spanned by temperature and chemical potential ({see {Refs.}~\cite{Thies:2003br} for analytic studies).}
{With our numerical approach, we} were able to reproduce the position of the phase boundary of the inhomogeneous phase at the level of 0.1$\%$.
\begin{figure}[t]
\includegraphics[width=1\columnwidth]{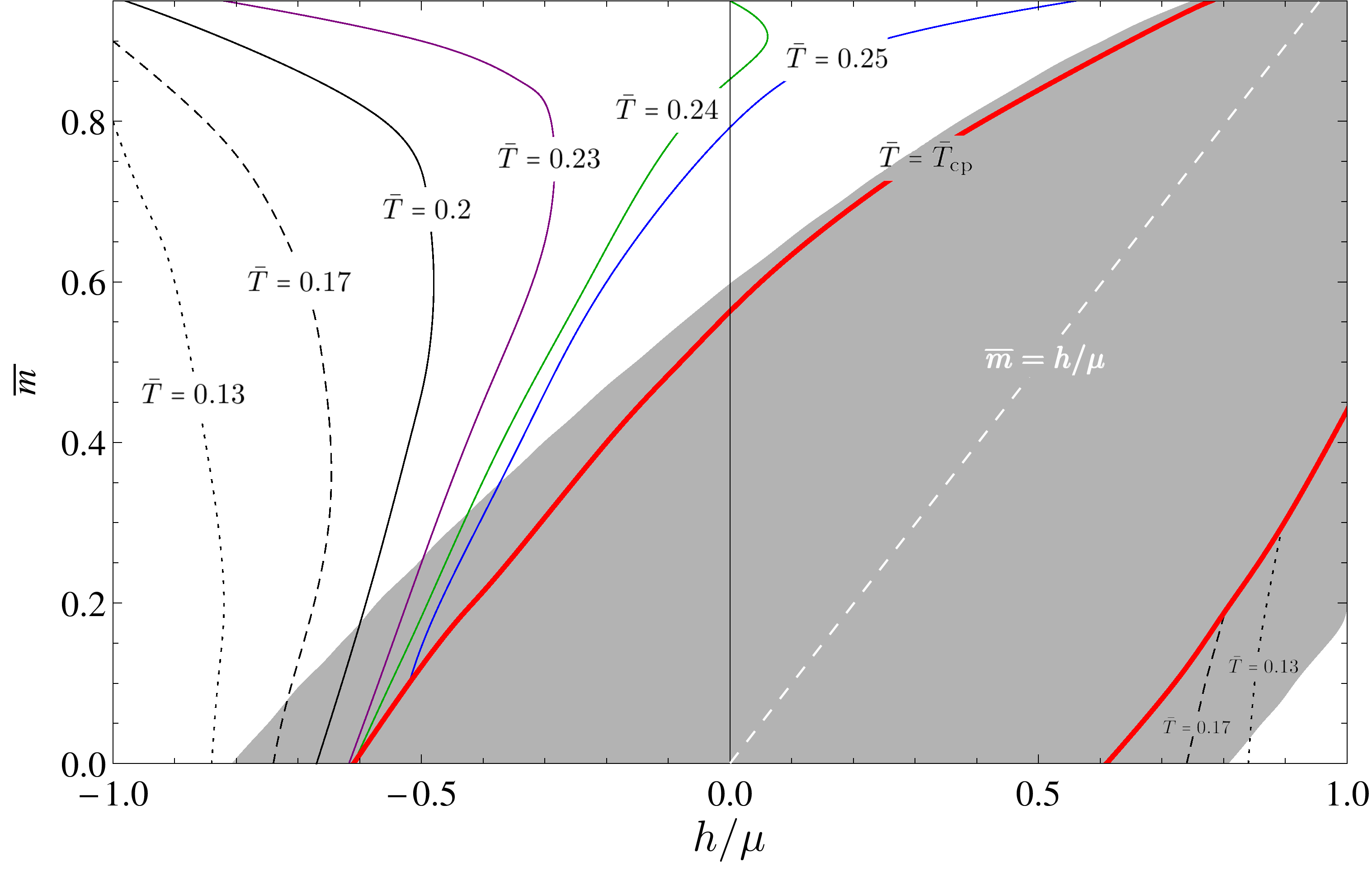}
\caption{\label{fig:p2d} 
(Color online) FFLO phase boundaries for {various dimensionless temperatures $\bar{T}=T/\mu$ in the $(\bar{m},h)$} plane. The red line presents the 
position of the critical points~$\bar{T}_{\rm cp}$ at which the FFLO phase boundary merges into the BCS phase boundary. For comparison, the gray-shaded
area gives the size of the BCS phase at $T=0$ as obtained from a standard mean-field study, i.e. without taking
into account the possibility of an inhomogeneous ground state. 
{Our results suggest that} the latter studies overestimate the size of the homogeneous phase in the~$(\bar{m},h)$ plane.} 
\end{figure}

{The phase diagram of the strongly-interacting one-dimensional Fermi gas in the $(T,\bar{m},h)$ space 
for $k_{\rm F}a_{\rm s}=\sqrt{2m_{\rm r}\mu/g^2}=\sqrt{\mu/(2g^2)}= 1/(\sqrt{2}\pi)$ is shown
in Fig.~\ref{fig:p3d}.}
To obtain this diagram, we used $n_{\rm max}=2$ in the expansion of the effective action and {up to}~$l_{\max}=5$ for the 
expansion of the field~$\varphi$. We have checked that our results for the various phase boundaries are not altered by 
increasing~$n_{\text{max}}$, at least to the accuracy of our calculations. The values of observables other than the phase 
boundaries ({such as the fermion gap}) {may, however, be} significantly altered when~$n_{\text{max}}$ is {increased, at least for~$n_{\text{max}}\leq 4$.}

The (orange-brown) `dome' in Fig.~\ref{fig:p3d} depicts the standard BCS phase with a homogeneous ground state. In this phase, the U($1$) 
symmetry is spontaneously broken, which is associated with superfluid behavior. The 
appearance of such a conventional BCS phase is not unexpected. In fact, at $h/\mu=\bar{m}$, the Fermi momenta associated with the up and 
down fermions coincide, which favors BCS-type pairing. Increasing the temperature at fixed~$\bar{m}$ and~$h$, within the BCS phase, 
a critical temperature exists above which the U($1$) symmetry is restored and superfluid behavior is lost. Specifically, 
we find~$T_{\rm cr}/\mu=0.55$ for $\bar{m}=h=0$. The phase transition between the BCS phase and the U($1$)-symmetric phase (normal phase) 
is of {second order. 

Note that the theory} is \emph{not} invariant under~$h\to -h$ for given fixed
temperature and a finite mass-imbalance~$\bar{m}$; it is only invariant under a simultaneous change of the sign of~$\bar{m}$ and~$h$.
From an experimental point of view, this implies that the finite-temperature phase diagram of a spin-imbalanced ${}^{6}\text{Li}\text{-}{}^{40}\text{Ka}$-
mixture~($\bar{m}\approx 0.74$) is substantially different from that of a mass-balanced gas (see also Fig.~\ref{fig:p2d}). In fact, our study suggests 
that, depending on $a_{\rm s}$, a large part of {the (finite-temperature) phase diagram} is occupied by the inhomogeneous 
phase for the ${}^{6}\text{Li}\text{-}{}^{40}\text{Ka}$-mixture, see also below for a discussion of the dependence on~$a_{\rm s}=1/g$.
\begin{figure*}[t]
\includegraphics[width=0.581\columnwidth,angle=-90]{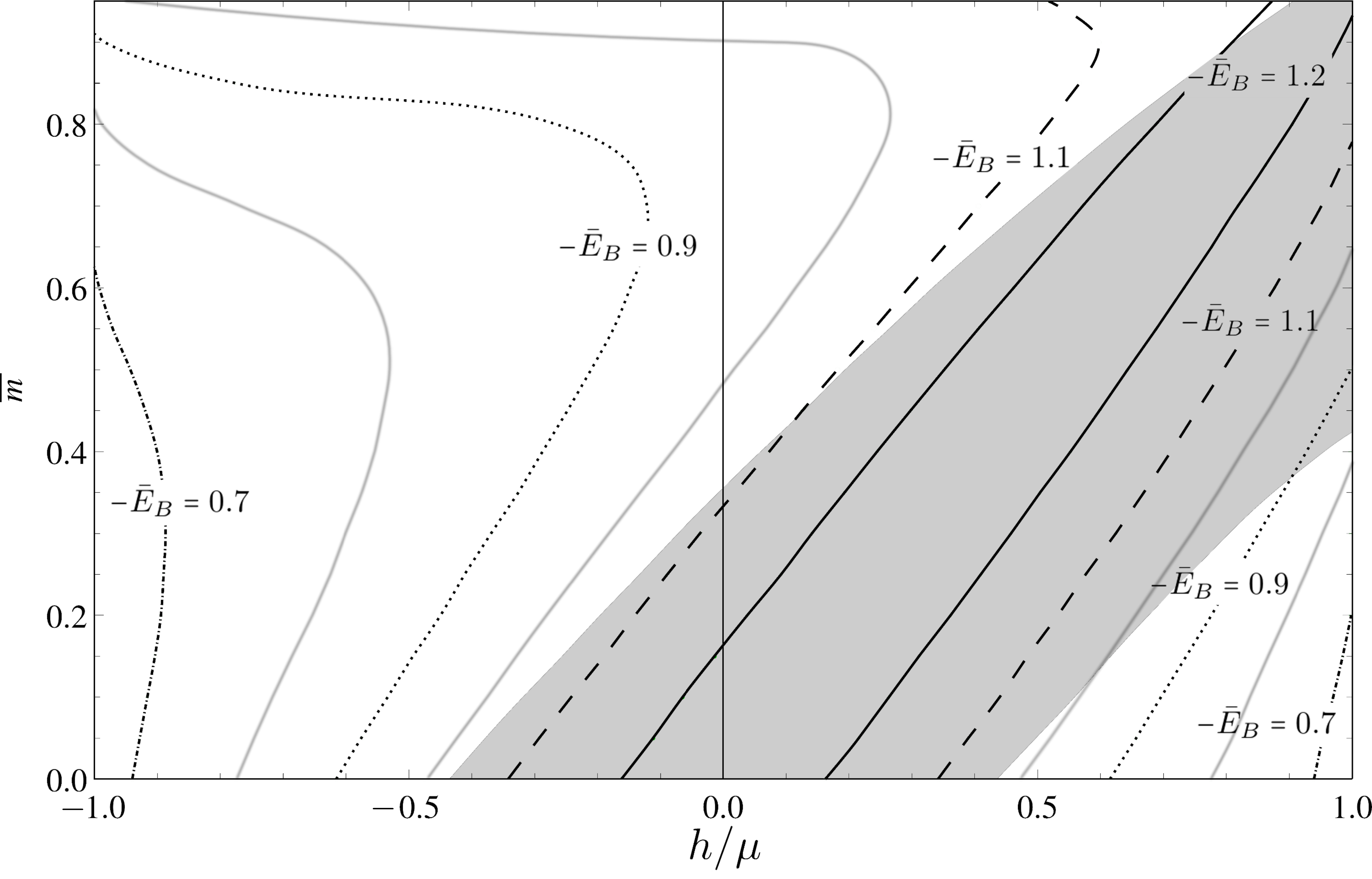}\hspace*{1.1cm}
\includegraphics[width=0.581\columnwidth,angle=-90]{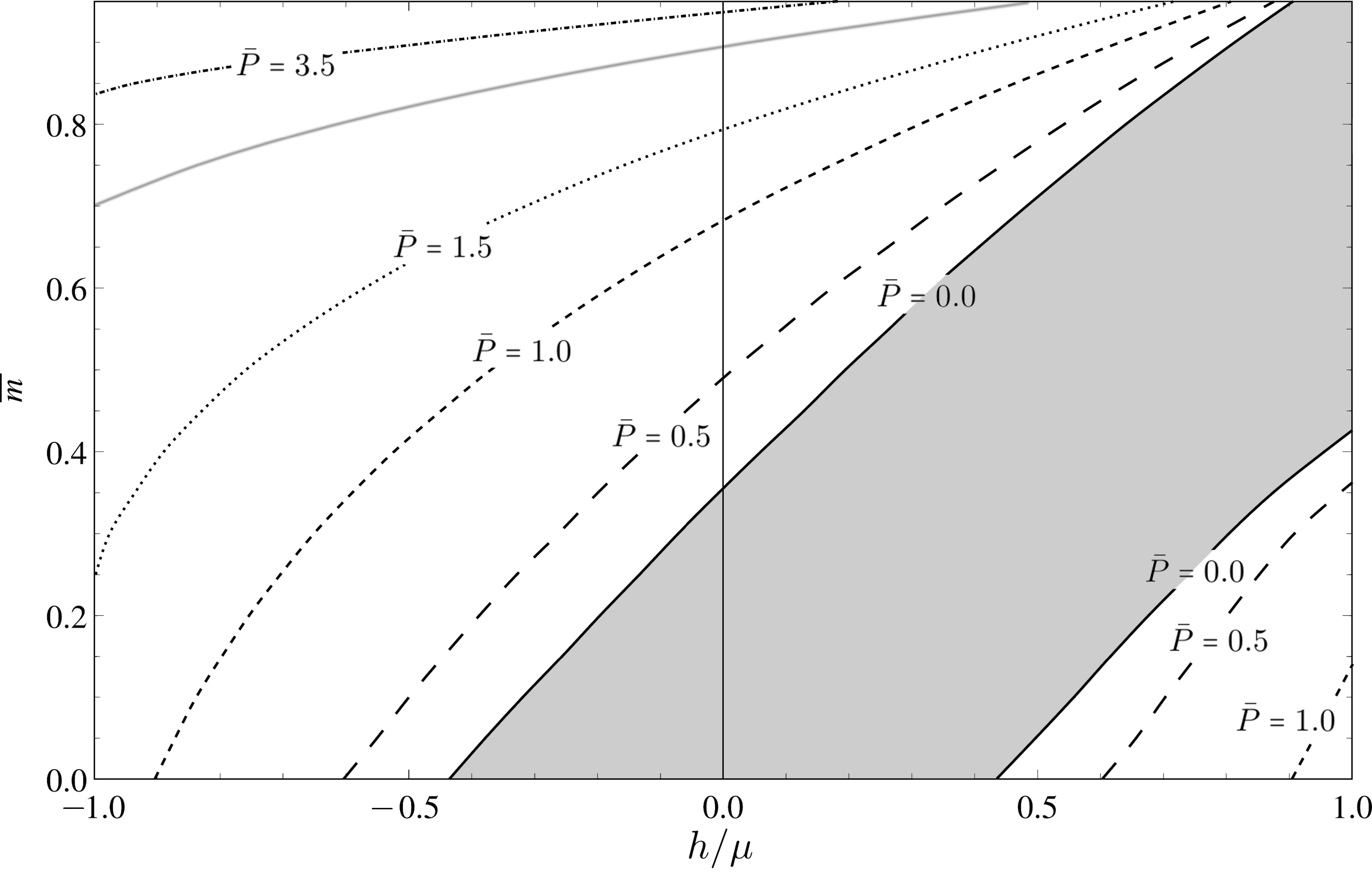}
\caption{\label{fig:2bp} 
Left panel: {Dimensionless binding energy~$\bar{E}_{\rm B}=E_{\rm B}/\mu$ of} the {lowest-lying} two-fermion bound state in the plane spanned by the mass imbalance~$\bar{m}$ and
the spin polarization parameter~$h$. The gray-shaded area has been included to guide the eye and depicts the regime in which the total momentum~$P$ of the bound state 
is zero. Right panel: Dimensionless center-of-mass {momentum $\bar{P}=P/\sqrt{\mu}$ of the} bound state in the $(\bar{m},h)$ plane. 
}
\end{figure*}

At sufficiently low temperatures, our results indicate that the system undergoes a phase transition from the homogeneous BCS phase
to a FFLO-type phase characterized by an inhomogeneous ground state (see also below). The latter is portrayed by 
the greenish-colored areas to the left and to the right of the BCS `dome'  in Fig.~\ref{fig:p3d}. Here, the ground state is also associated with superfluid 
behavior, but translational invariance is broken spontaneously as well. {For the ground-state configurations within} this phase, we find from our numerical 
analysis that~$\alpha^{\rm gs}_l=0$ for even $l\geq 0$, 
and~$\alpha^{\rm gs}_l\neq 0$ otherwise. Moreover, we observe that $\alpha^{\rm gs}_1$ is at least one order of magnitude larger 
than~$\alpha^{\rm gs}_3$ which is again at least one order of magnitude larger than~$\alpha^{\rm gs}_5$. This justifies the truncation of the sum in 
Eq.~\eqref{eq:phiexp} and suggests that the spatial structure of the inhomogeneity is dominated by the coefficient~$\alpha^{\rm gs}_1$.
Qualitatively, the existence of such a FFLO-type phase for a given temperature can be traced back to the fact that the difference between the Fermi 
momenta of the two fermion species increases away from the line defined by~$h/\mu=\bar{m}$. Far away from this line, pairing of two fermions 
with opposite spins is only possible if the mismatch between the associated Fermi momenta can be compensated by a finite center-of-mass momentum 
of the pair. In the limit of vanishing temperature, we then observe that there is no (quantum) phase transition from the FFLO phase to the normal phase, 
even for large mass- and/or spin imbalances. For fixed~$\bar{m}$ and~$h$ within the FFLO phase, on the other hand, we observe that the 
system undergoes a second order phase transition when the temperature is increased. 

The homogeneous, FFLO and normal phases coexist on a line that we define as~$T=T_{\rm cp}(\bar{m},h)$, and is depicted in Fig.~\ref{fig:p2d}. 
{Interestingly, we find that the phase transition from the BCS-type to the FFLO-type phase is of first order,}
at least for the considered truncations of the effective action,~$n_{\max}=2,3,4$, 
(see white-shaded area in Fig.~\ref{fig:p3d}). Note that the corresponding phase transition line in the GN model was found to be of second 
order~\cite{Thies:2003br}. {However, an exact determination of the order of this transition is beyond the scope of the present work.}

Let us now consider finite temperatures $T<T_{\rm cp}$. 
For a given~$\bar{m}$, a second-order phase transition from
the FFLO phase to the normal phase occurs for sufficiently large values of the spin-imbalance parameter~$h$.
In Fig.~\ref{fig:p2d}, we show the shape of this phase transition line in the $(\bar{m},h)$ plane for various different temperatures. We find that
{the corresponding transition line develops an intriguing back-bending shape when the temperature is lowered.} 


{The basic mechanism underlying this characteristic back-bending shape of the  
transition line} between the FFLO and the normal phase can be most easily understood by considering the two-body problem
in the presence of two Fermi surfaces as described by the following {Schr\"odinger equation:
\be
\label{eq:sg}
\Big[ \sum_{\sigma=\uparrow,\downarrow}\epsilon_{\sigma}(\partial_{x_{\sigma}}) - g\delta(x_\uparrow-x_\downarrow)
+ E_{\rm B}\Big]\Psi(x_\uparrow,x_\downarrow)=0\,.
\ee
Here,~$\Psi$ is the wave-function of the bound state.} The operator~$\epsilon_{\sigma}$ is defined 
as~$\epsilon_{\sigma}(\partial_{x_{\sigma}})=|-(2m_{\sigma})^{-1}\partial_{x_{\sigma}}^2-\epsilon_{\rm F,\sigma}|$ with~$\epsilon_{\rm F,\sigma}$ $(\sigma=\uparrow,\downarrow)$ 
being the Fermi energy of the up- and down-fermions, respectively. 
Along the lines of the two-body problem in three dimensions, see, e.g., Ref.~\cite{Pitaevskii}, the solution of our one-dimensional two-body problem can in principle 
also be given in closed form. 
For our purposes, rather the (binding) energy of the lowest-lying bound-state, which is  obtained from a minimization of the energy~$E_{\rm B}$ with respect to the  
total momentum~$P$, is of particular interest.
For~$k_{\rm F}a_{\rm s} = 1/(\sqrt{2}\pi)$, the energy of this state and its total momentum are  
given in Fig.~\ref{fig:2bp}.

From our results of the two-body problem, we reveal, analogous to BCS theory, 
an intriguing relation between the bound-state properties of the two-body problem and the phase 
structure of the many-body problem
in {the $(T,\bar{m},h)$ space.} Although condensation in a many-body system is clearly a collective phenomenon, the condensation pattern in the phase diagram appears to be dictated {largely}
by the underlying two-body {physics}. Along the line~$h/\mu=\bar{m}$, we find that the binding energy is minimal and the center-of-mass momentum of the lowest-lying state is zero. This regime translates
into the homogeneous BCS phase in the many-body problem.  
{Increasing~$h/\mu$ starting from a given~$h/\mu=\bar{m}$, we} 
{observe that~$|E_{\rm B}|$ decreases} and the lowest-lying state eventually assumes 
a finite center-of-mass momentum~$P$. 
Unless a finite~$P$ is energetically favored, we do not expect to find
a FFLO-type phase in the many-body phase diagram. This is indeed the case. Moreover, we find that {the characteristic back-bending shape of the FFLO transition line 
in the many-body phase
diagram (see Fig.~\ref{fig:p2d}) appears to be dictated by the shape for the lines of constant binding} energy shown in Fig.~\ref{fig:2bp}. For increasing temperature, 
it is in fact reasonable to expect that the system is first pushed into the normal phase in those domains of {the $(\bar{m},h)$ plane} in which the absolute value of the 
binding energy is minimal. Although the exact values of the phase transition temperature of the many-body problem 
cannot be deduced from the two-body problem, our study confirms that the general phase structure is directly related to the underlying two-body problem. {In this respect, it is worth to add that}
the size of the various regimes with finite and vanishing center-of-mass momentum depends on the actual value of the coupling~$g$. {To be specific, we find} that the regime with~$P=0$ increases
for increasing~$g$ and shrinks for decreasing~$g$. 

{\it Summary -} We have studied the finite-temperature phase diagram of mass- and spin-imbalanced Fermi gases
in one spatial dimension. In our analysis we focused on the strongly coupled regime and allowed for spatially varying configurations
of {the order-parameter field~$\varphi$.} The latter enabled us to discern the appearance of inhomogeneous phases, which 
appear in a considerable region of parameter space. In particular, we found an intriguing back-bending structure in the finite-temperature phase diagram, 
which can be explained by analyzing the two-body problem in the presence of Fermi surfaces. {The robustness of this structure in the presence of a finite
transverse coupling between the tubes is an important question that can be addressed in the future with our approach. In particular,}
our analysis of the general phase structure is not limited to {the strict one-dimensional} case but 
may also be applied to higher-dimensional Fermi gases; it may therefore contribute to better our understanding of strong fermion dynamics {also}
beyond ultracold quantum gases. 
 

{{\it Acknowledgements -}} J.B. would like to thank B.~Drossel and F. Karbstein for useful discussions. 
J.B. and D.R. acknowledge support by the DFG under Grant BR 4005/2-1. Moreover, J.B. acknowledges support
by HIC for FAIR within the LOEWE program of the State of Hesse. J.E.D. acknowledges funding from the U.S. National Science Foundation,
under grant No. PHY1306520.

\bibliographystyle{h-physrev3}

\end{document}